\documentclass[sn-nature,pdflatex]{sn-jnl}

\usepackage{graphicx}%
\usepackage{multirow}%
\usepackage{amsmath,amssymb,amsfonts}%
\usepackage{amsthm}%
\usepackage{chngpage}
\usepackage{mathrsfs}%
\usepackage[title]{appendix}%
\usepackage{xcolor}%
\usepackage{textcomp}%
\usepackage{manyfoot}%
\usepackage{booktabs}%
\usepackage{algorithm}%
\usepackage{algorithmicx}%
\usepackage{algpseudocode}%
\usepackage{listings}%
\usepackage{layouts}%
\usepackage{verbatim}%
\usepackage{placeins}%
\usepackage{hyperref}%
\usepackage{nicefrac}%
\usepackage{tabularx}%
\usepackage{caption}%
\usepackage{makecell}%
\usepackage{siunitx}
\DeclareSIUnit{\crate}{C}
\DeclareSIUnit{\ah}{\ampere\hour}
\DeclareSIUnit{\kwh}{\kilo\watt\hour}
\DeclareSIUnit{\wh}{\watt\hour}
\DeclareSIUnit{\kw}{\kilo\watt}
\usepackage{acro} 
\usepackage{array}
\newcolumntype{L}[1]{>{\raggedright\let\newline\\\arraybackslash\hspace{0pt}}m{#1}}
\newcolumntype{C}[1]{>{\centering\let\newline\\\arraybackslash\hspace{0pt}}m{#1}}
\newcolumntype{R}[1]{>{\raggedleft\let\newline\\\arraybackslash\hspace{0pt}}m{#1}}
\usepackage[final, commandnameprefix=always]{changes}
\setdeletedmarkup{\textcolor{red}{#1}}
\DeclareAcronym{AC}{
	short = AC,
	long = alternating current,
}
\DeclareAcronym{BMS}{
	short = BMS,
	long = battery management system,
}
\DeclareAcronym{CAN}{
	short = CAN,
	long = controller area network,
}
\DeclareAcronym{CP}{
	short=CP,
	long=constant power,
}
\DeclareAcronym{DC}{
	short=DC,
	long=direct current,
}
\DeclareAcronym{DV}{
	short=DV,
	long=differential voltage,
}
\DeclareAcronym{ECM}{
	short = ECM,
	long = equivalent circuit model,
	long-plural-form = equivalent circuit models,
}
\DeclareAcronym{EU}{
	short=EU,
	long={European Union},
}
\DeclareAcronym{EV}{
	short=EV,
	long={electric vehicle},
}
\DeclareAcronym{HPPC}{
	short = HPPC,
	long = hybrid pulse power characterization,
}
\DeclareAcronym{HV}{
	short=HV,
	long={high voltage},
}
\DeclareAcronym{HVAC}{
	short=HVAC,
	long={heating, ventilation, and air conditioning},
}
\DeclareAcronym{ICA}{
	short=ICA,
	long={incremental capacity analysis},
}
\DeclareAcronym{IC}{
	short=IC,
	long={incremental capacity},
}
\DeclareAcronym{ICEV}{
	short=ICEV,
	long={internal combustion engine vehicle},
}
\DeclareAcronym{ID}{
	short=ID,
	long=identifier,
}
\DeclareAcronym{LAM}{
	short=LAM,
	long={loss of active material},
}
\DeclareAcronym{LFP}{
	short = LFP,
	long = lithium iron phosphate,
}
\DeclareAcronym{LIB}{
	short = LIB,
	long = lithium-ion battery,
	long-plural-form = lithium-ion batteries,
}
\DeclareAcronym{LLI}{
	short=LLI,
	long={loss of lithium inventory},
}
\DeclareAcronym{NE}{
	short=NE,
	long={negative electrode},
}
\DeclareAcronym{NMC}{
	short = NMC,
	long = nickel cobalt manganese oxide,
}
\DeclareAcronym{OBD}{
	short=OBD,
	long={onboard diagnosis},
}
\DeclareAcronym{OCPP}{
	short = OCPP,
	long = open charge point protocol,
}
\DeclareAcronym{OCV}{
	short = OCV,
	long = open circuit voltage,
}
\DeclareAcronym{OEM}{
	short = OEM,
	long = original equipment manufacturer,
}
\DeclareAcronym{OTA}{
	short = OTA,
	long = over-the-air,
}
\DeclareAcronym{PE}{
	short=PE,
	long={positive electrode},
}
\DeclareAcronym{SEI}{
	short=SEI,
	long={solid electrolyte interphase},
}
\DeclareAcronym{SOC}{
	short = SOC,
	long = state of charge,
}
\DeclareAcronym{SOCE}{
	short = SOCE,
	long = state of certified energy,
}
\DeclareAcronym{SOCR}{
	short = SOCR,
	long = state of certified range,
}
\DeclareAcronym{SOH}{
	short = SOH,
	long = state of health,
}
\DeclareAcronym{UBE}{
	short=UBE,
	long={usable battery energy},
}
\DeclareAcronym{UDS}{
	short=UDS,
	long={uniﬁed diagnostic services},
}
\DeclareAcronym{UI}{
	short = UI,
	long = user interface,
}
\DeclareAcronym{UN}{
	short = UN,
	long = United Nations,
}
\DeclareAcronym{VW}{
	short = VW,
	long = Volkswagen,
}
\DeclareAcronym{WLTC}{
	short = WLTC,
	long = Worldwide harmonized Light Duty Test Cycle,
}
\DeclareAcronym{WLTP}{
	short = WLTP,
	long = Worldwide Harmonized Light Vehicles Test Procedure,
}

\begin{document}

\title[Article Title]{Why we need a standardized state of health definition for electric vehicle battery packs -- a proposal for energy- and capacity-based metrics}

\author*[1]{\fnm{Philip} \sur{Bilfinger}}\email{philip.bilfinger@tum.de}
\author[1]{\fnm{Markus} \sur{Schreiber}}\email{markus.schreiber@tum.de}
\author[1]{\fnm{Philipp} \sur{Rosner}}\email{philipp.rosner@tum.de}
\author[1]{\fnm{Kareem} \sur{Abo Gamra}}\email{kareem.abo-gamra@tum.de}
\author[1]{\fnm{Jan} \sur{Sch{\"o}berl}}\email{jan.schoeberl@tum.de}
\author[1]{\fnm{Cristina} \sur{Grosu}}\email{cristina.grosu@tum.de}
\author[1]{\fnm{Markus} \sur{Lienkamp}}\email{lienkamp@tum.de}

\affil[1]{\orgdiv{Institute of Automotive Technology (FTM)}, \orgname{Technical University of Munich (TUM)}, \orgaddress{\street{Boltzmannstr. 15}, \city{Garching}, \postcode{85748}, \state{Bavaria}, \country{Germany}}}


\abstract{Range and performance are key customer-relevant properties of electric vehicles. Both degrade over time due to battery aging, thus impacting business decisions throughout a vehicle's lifecycle, such as efficient utilization and asset valuation. For practical assessment, aging is often simplified into a single figure of merit -- the state of health -- typically defined by the battery pack’s remaining capacity or energy. However, no standardized method for measuring the state of health at the vehicle level has been established, leaving both academia and industry without a clear consensus. Ultimately, standardization is crucial to increase transparency and build confidence in the long-term reliability of electric vehicles' battery packs.

In this article, we propose a standard measurement procedure for assessing the capacity- and energy-based state of health, leveraging onboard charging to enable reproducibility and scalability. Additionally, we demonstrate how differential voltage analysis can provide deeper insights into battery aging at the vehicle level.}




\maketitle
\section*{Introduction}
\section*{}
\label{ch:intro}

\Acp{EV} have emerged as the breakthrough solution to decarbonize and electrify the transportation sector, capable of significantly reducing greenhouse gas emissions~\cite{Woody.2023}. Enablers for the acceptance of electric mobility and exponential growth in \ac{EV} sales are advancements and innovations in their battery packs, which are currently \chreplaced{mostly}{almost exclusively} assembled with \acp{LIB}~\cite{Duhnen.2020}. These battery packs must satisfy the customer's required achievable range and power capabilities, embodying central properties for purchasing an \ac{EV}~\cite{JohnKiserandGrahamGordon.2023}. 
However, these capabilities degrade throughout the \ac{EV}'s usage life due to a complex and coupled interplay of reversible and irreversible aging mechanisms~\cite{Edge.2021}. Typically, battery aging is \chadded{generally} expressed by means of the \ac{SOH} that aggregates degradation in the remaining capacity or energy, expressed as a scalar value in reference to pristine conditions~\cite{Demirci.2024}. On the vehicle level, this offers a \chreplaced{simple}{facile} approach directly correlating to the remaining achievable range of an \ac{EV}~\cite{WASSILIADIS2022100167}. However, to our knowledge, neither industry nor academia has a consensus on a vehicle-level \ac{SOH} definition or a standard procedure for its measurement, lacking standardization and a common understanding of how aging shall be characterized for \acp{EV}. Furthermore, only considering the overall capacity- or energy loss superficially reflects battery aging, i.e., \acp{EV} with the same \ac{SOH} might have degraded through different aging mechanisms~\cite{Dubarry.2020}.

Hence, with increasing efforts to establish a holistic and circular battery ecosystem, it is vital to precisely express battery degradation to achieve an accurate \ac{SOH} representation~\cite{Hu.2020}. This allows stakeholders along the entire \ac{EV} value chain to operate the battery within ideal parameters and make informed decisions regarding maintenance, use in second-life applications, or recycling, thus maximizing resource efficiency and reducing costs~\cite{Harper.2019}. Furthermore, an accurate depiction in the \ac{SOH} provides transparency regarding insurance premiums or liability decisions in warranty claims~\cite{Sulzer.2021}. It also lays the foundation for assessing the residual value of secondhand vehicles~\cite{Schaltz.2021}. In summary, a transparent, reproducible standard measurement is necessary for a vehicle-level \ac{SOH} definition that covers the overall capacity/energy loss and enables deeper insights into the degradation condition of the battery pack. 

This article emphasizes the importance of vehicle-level \ac{SOH} standardization and proposes a transparent and reproducible standard measurement for capacity- and energy-based metrics derived from electrochemical properties of \acp{LIB} and designed for in-vehicle utilization. The overall aim is to close the gap between the necessity of a vehicle-level \ac{SOH} definition and the uncertainty of its determination, which is necessary for accepting battery electric vehicles as a suitable long-term alternative to conventional vehicles.

\section*{Results}
\subsection*{Regulation, standardization, and the EU battery passport}
\label{ch:EU}

The importance of \ac{SOH} standardization has reached the political arena as numerous political institutions have imposed and regularly expanded regulations to guarantee sustainability, safety, and quality standards for products containing batteries, with recent extensions to \acp{EV}~\cite{Junior.2024}.
Most notable is \ac{EU} Regulation 2023/1542, making a digital battery passport mandatory for new \acp{EV} entering service by 2027~\cite{EuropeanParliamentandCouncil.2023}.
The overall goal is to define technical standards, among which the \ac{SOH} is explicitly named as an essential performance metric of batteries, aiming at a more sustainable battery value chain. However, its articles concerning the \ac{SOH} lack technical detail and merely state that it must be given as an energy-based value for electric vehicles, named the \ac{SOCE}. Additionally, a rated capacity must be given in the vehicle registration documents, which is not explicitly appointed as a reference for the \ac{SOH} but can serve as one.
Ultimately, the regulation misses a definition of standard measurement and reference for obtaining the \ac{SOH} or \ac{SOCE}, which substantially impacts the values obtained.
This has also been criticized by the battery pass consortium~\cite{batterypass}, a publicly funded initiative uniting strategic partners in the field of \acp{EV}, which demands regulatory clarity.

Solely \ac{UN} Global Technical Regulation No. 22~\cite{UNGTR22}, incorporated into the EURO 7~\cite{EuropeanCommission.20240424} to regulate emissions and battery durability of motor vehicles, addresses the necessity and challenges for vehicle-level health metrics.
The regulation advocates for "in-use verification" via the tailored \ac{EV} driving cycle \ac{WLTC} through which the \ac{SOCE} and \ac{SOCR} are determined, but not the remaining capacity, which is, however, necessary for the previously mentioned \ac{EU} regulation.

International standards are equally unsuited, e.g., ISO~12405-4~\cite{InternationalOrganizationforStandardization.201807} vaguely proposes standard (dis)charging cycles for battery packs of \acp{EV} relying on the battery "supplier's recommendations," which are not publicly available.

The goal of \ac{SOH} standardization for \acp{EV} is definite from a political point of view. However, the labyrinth of regulation and standardization lacks a clear definition of what constitutes a vehicle level \ac{SOH}, to which reference it is based, and how it shall be measured.

\subsection*{Viable approaches for a vehicle-level battery state of health standard measurement}
\label{ch:SOTA}

The general methodologies for defining a standard measurement for an \ac{EV}'s \ac{SOH} are reviewed in this section and can be classified into model-based and measurement-based approaches.

Model-based approaches estimate the \ac{SOH}, e.g., through electrochemical or equivalent circuit models, but inherit model inaccuracies and are either unsuitable for online application or imprecise~\cite{Yao.2021}. The \ac{SOH} read directly from the vehicle's \ac{BMS} is estimated by such models~\cite{Demirci.2024}. However, this \ac{SOH} is supplied by the \ac{OEM}, which can have an incentive to manipulation so that it does not necessarily coincide with actual battery aging. Therefore, exclusive reliance on model-based approaches is not recommended.

\begin{figure}[h]
	\centering
	\makebox[0pt]{\includegraphics[width=14cm]{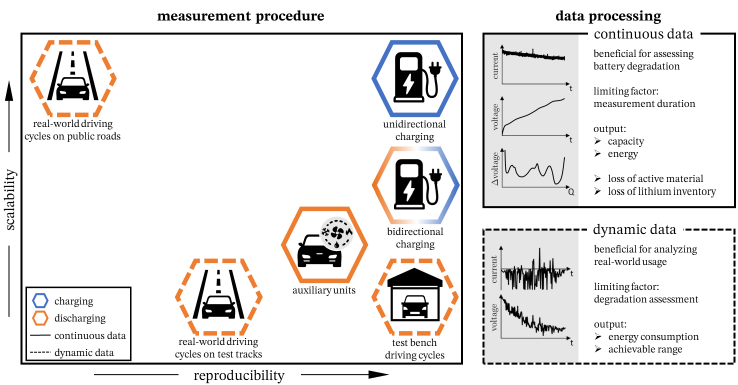}}
	\caption{Qualitative evaluation of viable \ac{SOH} standard measurements at the vehicle level, focusing on scalability for in-vehicle utilization and reproducibility of the measurement procedure. A standard measurement method based on unidirectional charging satisfies both criteria. The data obtained from \ac{EV} measurements can either be continuous or dynamic. The former is suited for deeper battery degradation analysis, and the latter for real-world consumption and range analysis.}
	\label{fig:abstract}
\end{figure}

Measurement-based approaches actively stress the \ac{EV}'s battery pack either by discharging or charging, which is further discussed in terms of their suitability for standard measurement and data processing, with an overview in \autoref{fig:abstract}.

Discharging the battery pack during driving cycles mimics an \ac{EV}'s actual operation and use case, which can be performed on test benches (e.g., chassis dynamometer) or on the road. The former ensures reproducibility but is cumbersome, time-consuming, and expensive, impairing scalability. Real-world driving profiles on the road are an application advantage but not reproducible due to fluctuating external influences. Furthermore, due to the dynamic loads, insights into the batteries' behavior and degradation states are difficult to obtain. Therefore, driving profiles, neither on a test bench nor on (public) roads, are suitable as a standard measurement for the \ac{SOH} but are important for actual real-world energy consumption and range estimations of \acp{EV}.

Continuous discharging of the battery by auxiliary units is possible for a reproducible measurement but is time-consuming and less applicable in operation as energy is wasted through heat dissipation, and recharging is necessary for the following use. A bidirectional charging standard can provide an opportunity for a controlled discharging procedure and automated recharging without significant energy loss, but it has yet to be widely adopted~\cite{Gutwald.2024}.

A scalable and reproducible approach relies on onboard charging, as stressing the battery pack takes place in a controlled manner defined by standardized protocols~(e.g., ISO~15118 or IEC~61851). Hereby, thermodynamic changes in the voltage signal can be captured for further degradation assessment, e.g., by \ac{DV} analysis, which can provide detailed information on where the overall capacity/energy loss occurs~\cite{Bilfinger.2024}. The measurement can conveniently be integrated into vehicle operation to monitor the \ac{SOH} throughout the vehicle's lifetime, as \acp{EV} must regularly be charged, and the availability of chargers has increased throughout the years, especially wallboxes for homeowners~\cite{IEA2024GlobalEV}. Hereby, using \ac{AC} or \ac{DC} chargers is arbitrary when sampling data from the \ac{EV}'s \ac{BMS}, as the charging power is ultimately transformed into the battery pack's \ac{DC} system. A benefit of \ac{DC} charging is the capability of charger-side data acquisition due to necessary standardized communication norms~(e.g., ISO~15118), where the battery pack's voltage and current values are periodically transmitted to the charger's backend. Given these advantages, onboard charging is best suited for developing a \ac{SOH} standard measurement for \ac{EV} battery packs.

\subsection*{A standard charging measurement proposal}
\label{ch:standard_measurement}

\begin{figure}[h]
	\centering
	\includegraphics[width=88mm]{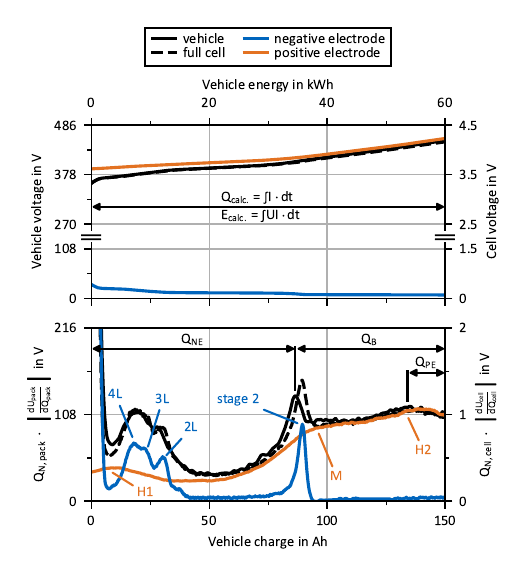}
	\caption{Charging measurement of a \ac{VW} ID.3 following the proposed standard measurement. a) Full and half-cell laboratory measurements from a  VW ID.3 pouch cell are aligned to the vehicle's voltage and capacity data, showing the transferability of cell-specific characteristics like the potential contribution of the positive and negative electrodes during (de)lithiation. b) \ac{DV} curves are calculated from the derivation of the voltage signal and enable deeper analysis of the battery's electrochemical condition. Changes in the characteristic capacities $Q_{NE}$, $Q_{PE}$, and $Q_{B}$ throughout aging enable the calculation of degradation modes to determine \ac{LAM} at the \ac{NE}, \ac{PE}, and \ac{LLI}~\cite{Keil.2017}. \ac{DV} analysis can also be utilized on the vehicle level, but limitations are to be considered, e.g., cell-to-cell variations in the pack (see differences at stage 2)~\cite{Bilfinger.2024}.}
	\label{fig:nmc811}
\end{figure}

Any standard measurement must be unambiguously defined for transparency and reproducibility. For a standard charging measurement, shown in \autoref{fig:nmc811}a for a \ac{VW} ID.3, this implies fixed conditions for the charging procedure, including the charging rate and cut-off constraints, the ambient temperature, and the preparation setup~\cite{Barai.2019}. 

The maximum charging rate must meet a trade-off between the overall measurement duration and the containment of overpotentials that stem from ohmic and kinetic resistances in the batteries~\cite{Kim.2022}. The latter limits the transferable capacity/energy, which are, however, the quantities of interest~\cite{Barai.2019}. Therefore, a measurement duration of at least $t=$ \SI{15}{\hour} is proposed to contain the influence of overpotentials. The measurement time seems lengthy but yields the most accurate results, which is further discussed in \autoref{ch:Dis}.
The maximum charging power can be determined for each \ac{EV} by the $E-rate$, which is the inverse of the charging time $t$ normalized by the battery pack's net energy, as given in \autoref{eq:E-rate}.
\begin{align}
	E-rate = \frac{P_{max}}{E_N}<\frac{1}{t}\Rightarrow P_{max} < \frac{1}{t} \cdot E_N\label{eq:E-rate}
\end{align}
With typical battery pack energies up to \SI{120}{\kilo\watt\hour}~\cite{EV_statistics}, this results in a maximum charging power of \SI{8}{\kilo\watt}, which is below typical wallbox powers in the \ac{EU} of \SI{11}{\kilo\watt} enabling simple utilization. For charging, we propose a \ac{CP} protocol, as it is common for \ac{EV} charging due to energy conservation with the grid.

Furthermore, the minimum measuring window, i.e., the cut-off conditions, must be defined either by a \ac{SOC} window or voltage range. Fixing a voltage range is unambiguous, as the voltage is a directly measurable quantity, unlike the \ac{SOC}, which can underlie estimation errors and is prone to manipulation~\cite{Danko.2019}. However, the voltage is less accessible and currently must be reverse-engineered from \ac{CAN} data so that simple access to the voltage must be granted, e.g., through the \ac{OBD}-II interface~\cite{rosenberger2024quantifying}. Nonetheless, a fixed pack-level voltage window should be used for reproducible and comparable measurements, which must be declared in the vehicle's registration documents. The voltage range should hereby be as broad as necessary to capture characteristic changes in the voltage signature but as narrow as possible with the cut-off voltages reachable in the typical operation window of an \ac{EV} to ensure applicability. Some authors seek applicability by correcting a partial charging measurement with the \ac{SOC} range of the measurement, i.e., $Q = \nicefrac{Q_{partial}}{\Delta SOC}$~\cite{Qi.2024,Farmann.2015}. However, a direct measurement without the estimated \ac{SOC} is more accurate.

\chdeleted{Additionally, a resting period prior to the measurement is recommended to allow potential gradients in the battery pack to fade~\cite{Rashid.2015}. The required duration depends on prior excitation, leading to an exponential overpotential decay. A practical rule of thumb can be derived from Fernando et al.~\cite{Fernando.2024} that consider the voltage settled when its change falls below \SI{1}{\milli\volt\per\second}. For application in \SI{400}{\volt} \acp{EV}, this equates to a \SI{0.4}{\volt} difference between samples or a change smaller than the voltage sensor's resolution. For higher pack voltages, this threshold is scaled accordingly. In this study, a 30-minute rest period was sufficient to reach a stable voltage below the lower cut-off.}

\chadded{Additionally, a rest period prior to the measurement is recommended to allow potential gradients within the battery pack to decay exponentially, with the required duration depending on the excitation prior to the measurement~\cite{Rashid.2015}. A practical rule of thumb can be derived from Fernando et al.~\cite{Fernando.2024} that consider the voltage settled when its rate of change falls below $\SI{1}{\milli\volt\per\second}\cdot n_s$, where $n_s$ is the number of  serially connected cells, or when the variation is smaller than the voltage sensor's resolution. For example, a 30-minute rest period was sufficient in this study to achieve a stable voltage. Generally, the necessary resting time will increase when the battery pack is stressed with higher loads prior to the measurement or at lower temperatures~\cite{Fernando.2024}. Furthermore, the relaxed battery pack should settle below the predefined lower voltage threshold so that the defined voltage window is captured entirely.}

Lastly, the behavior of \acp{LIB} is strongly influenced by the temperature, so cold or hot ambient conditions shall be avoided for reproducibility~\cite{Waldmann.2014}. Hence, spatially measured temperatures within the battery pack should lie at room temperature of \SI{20}{\celsius} in a window of $\pm$\SI{5}{\celsius} before the measurement, and charging in an air-conditioned workshop or garage is recommended~\cite{Schaltz.2019}. Naturally, the battery's temperature increases during the charging measurement, which can additionally be monitored to detect hot spots within the pack. However, excessive temperatures are typically contained by an \ac{EV}'s thermal management system, e.g., the temperatures of the specimens under test never exceeded \SI{30}{\celsius} after the termination of the charging session~\cite{rosenberger2024quantifying}. 

The battery pack's capacity $Q_{calc}$ and energy $E_{calc}$ are then calculated by time integration between the lower cut-off voltage $U_{low}$ until the upper cut-off voltage $U_{high}$ is reached by 

\begin{align}
	Q_{calc} &= \int^{t_{U_{high}}}_{t_{U_{low}}} I\ dt \label{eq:capacity}\\
	E_{calc} &= \int^{t_{U_{high}}}_{t_{U_{low}}} P\ dt = \int^{t_{U_{high}}}_{t_{U_{low}}} U\ I\ dt \approx U_{n} \cdot Q_{calc}.
	\label{eq:energy}
\end{align}

The energy $E_{calc}$ is application-oriented, reflecting the usable energy in power-coupled systems, e.g., with inverters and electric motors, \chadded{and characterizes aging more closely to the customer's perception, specifically the vehicle's range at a given energy consumption}. In contrast, the capacity $Q_{calc}$ plainly accounts for the charge transferred between the electrodes, independent of the voltage \chadded{and the effect of overpotentials}. Therefore, capacity provides a consistent basis for comparing changes in electrode capacities to determine underlying degradation modes~\cite{Bulow.2023}.

For simplicity, some authors calculate the energy $E_{calc}$ by multiplying the calculated capacity by the nominal voltage $U_{n}$ of the battery pack, which must be available in the vehicle's registration documents, but only suits as an approximation~\cite{Jossen.2021}.
\newpage
Regularly performing this standard measurement and referencing the measured capacity and energy to the nominal values~$Q_{N}$/$E_{N}$ results in the capacity-based $SOH_Q$ and energy-based $SOH_E$ by
\begin{align}
	SOH_Q &= \frac{Q_{calc}}{Q_N} \label{eq:SOH_Q}\\
	SOH_E &= \frac{E_{calc}}{E_N} \label{eq:SOH_E}
\end{align}

If available, the calculated capacity and energy can be referenced to an initial value~$Q_{0}$/$E_{0}$ reflecting an \ac{EV}'s individual $SOH_{Q/E}$. 

\subsection*{Battery pack diagnostics with differential voltage analysis}
\label{ch:DVA}

The proposed 15-hour measurement is suited for vehicle-level \ac{DV} analysis, as analyzed by Schmitt et al.~\cite{Schmitt.2023}, enabling deeper insights into battery degradation through the identification of chemistry-specific characteristics. Nonetheless, lower charging powers are favorable with the cost of a longer measurement, as features for \ac{DV} analysis tend to be more pronounced~\cite{Fly.2020}. Thus, the measurements in this article were conducted at approx. \SI{2}{\kilo \watt}. In \ac{DV} analysis, local maxima, shown in \autoref{fig:nmc811}b for a \ac{NMC}/graphite chemistry, indicate lithiation stages of the \ac{LIB}, which can mostly be unambiguously assigned to either the \ac{PE} or \ac{NE}~\cite{Bloom.2005}. Tracking these maxima enables the calculation of the electrode capacities $\text{Q}_{\text{NE/PE}}$ and their balancing~$\text{Q}_{\text{B}}$. These capacities change with aging, from which the degradation modes \ac{LAM} at the electrodes and \ac{LLI} can be determined, and aging can be further dissected~\cite{Wildfeuer.2022}. Note that available characteristic features depend on the cell chemistry, so the degradation assessment must be tailored to the underlying battery pack~\cite{Dubarry.2020}. Nonetheless, the general methodology is transferable, as shown by a previous publication, where vehicles with different battery pack setups are compared~\cite{Bilfinger.2024}. \chadded{Also, other aging diagnostic methods, such as \ac{ICA}, benefit from the proposed charging procedure and can be utilized on the vehicle level~\cite{Schaltz.2021}.}
Nonetheless, it is clear that battery diagnosis, e.g., through \ac{DV} analysis, is not necessary for obtaining a battery pack's \ac{SOH} but rather enables test facilities for a more profound analysis of the battery's aging state.

\subsection*{Application of the proposed method}
\label{ch:diag}

\begin{figure}[b]
	\centering
	\makebox[0pt]{\includegraphics[width=17cm]{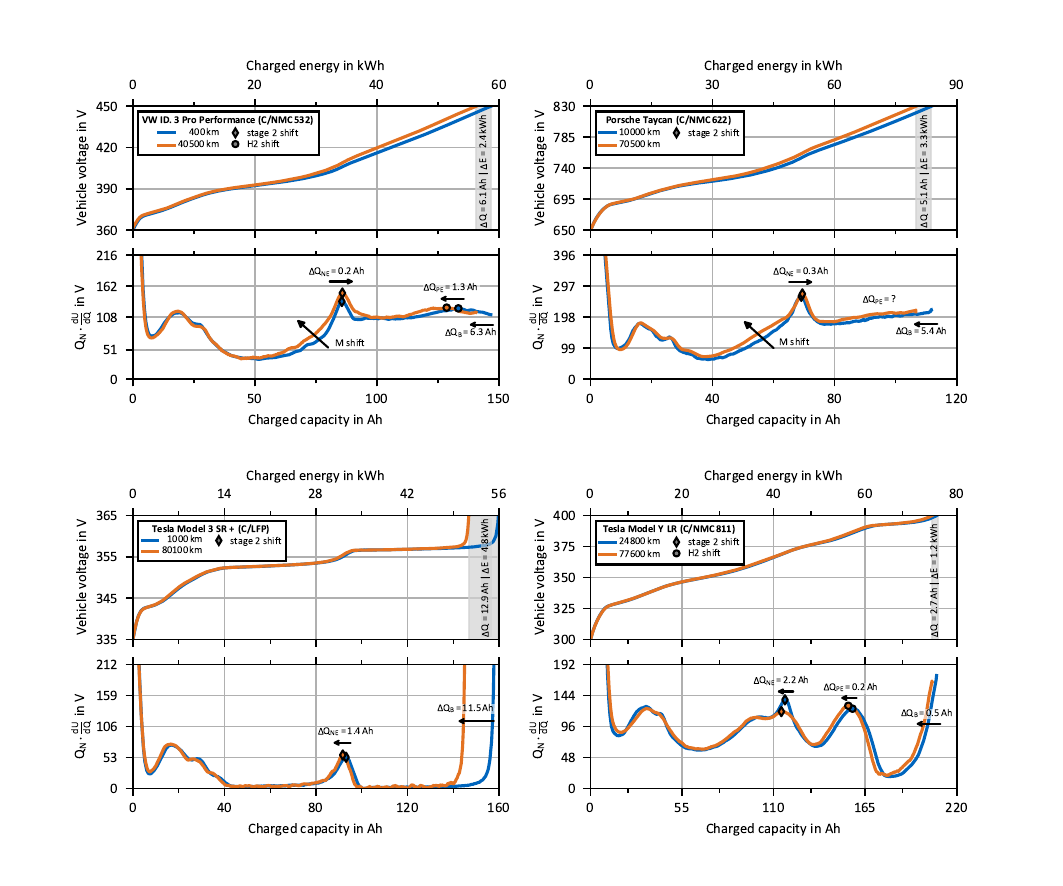}}
	\caption{Charging measurements of different vehicle models using the proposed standard charging measurement. a)-d) Voltage and \ac{DV} curves of the VW ID.3, Porsche Taycan, Tesla Model 3, and Tesla Model Y. Insights from the \ac{DV} curves show that \ac{LLI} is the primary degradation mode of these battery packs.}
	\label{fig:Def}
\end{figure}

The application and universality of the proposed method is shown in \autoref{fig:Def} on four state-of-the-art \acp{EV} --  the VW ID.3 Pro Performance, Porsche Taycan, Tesla Model 3 Standard Range Plus, and Tesla Model Y Long Range.

The curves show charging measurements at an early aging state and after utilization with more mileage covered. Evaluating the energy $E_{calc}$ and capacity $Q_{calc}$ between the specified voltage ranges given in \autoref{tab:vehicles} enables quantifying overall differences in the capacity/energy. Here, the \ac{SOH} metrics are calculated with the \ac{EV}s' nominal capacity and net energy, respectively, although these reference values were most likely not determined by the exact proposed standard measurement. Thus, it is explicitly emphasized that a reference is an essential value for \ac{SOH} comparability and must be available in the vehicle's registration documents with additional information available on how these values are determined.
Note that the Porsche Taycan and Tesla Model Y measurements were conducted with different vehicles of the same model type and battery pack specification, demonstrating comparability between specimens.

Deeper insights into the electrochemical aging of the battery packs are given through \ac{DV} analysis, where changes in the characteristic features $Q_{PE}$, $Q_{NE}$, and $Q_{B}$ are labeled. \ac{LAM} at the \ac{NE} is identified for the Tesla Model 3 and Tesla Model Y. \ac{LAM} at the \ac{PE} can only be identified for the \ac{VW} ID.3 and Tesla Model Y, as either there is no \ac{PE} feature~(e.g., for Tesla Model 3's \ac{LFP} cells) or the feature is not unambiguously detectable~(e.g., Porsche Taycan).
For all vehicles, \ac{LLI}, correlating with the change of the balancing feature $\text{Q}_{\text{B}}$, appears to be the predominant degradation mode at the time of the measurements and is mainly responsible for the loss of capacity/energy~\cite{Schreiber.2024}. This aligns well with aging studies on the cell level under real-world conditions, where aging mechanisms binding lithium-ions, such as \ac{SEI} formation, occur early in a battery's service life, directly resulting in \ac{LLI}~\cite{Geslin.2024}. \chadded{The quantitative results for the measured capacities, energies, and degradation modes are summarized in \mbox{Supplementary Table 1} in the supplementary information.}

\section*{Discussion}
\label{ch:Dis}

The proposed standard charging procedure aims to provide an unambiguous and reproducible measurement for determining an \ac{EV} battery pack's \ac{SOH}. 

The most apparent disadvantage of the proposed standard measurement lies in its duration. However, the measurement time is essential to ensure an accurate calculation of the \ac{SOH}, which is highly relevant to multiple stakeholders, given the battery pack’s critical role in vehicle utilization and asset valuation. Long idle periods -- such as overnight parking -- can be leveraged to conduct the measurement with minimal disruption. Moreover, the required measurement frequency remains low since battery aging generally occurs over several years. We therefore recommend performing the measurement once or twice per year. Exceptions apply in high usage cases, where more frequent measurements may be beneficial, or before the sale or purchase of a secondhand \ac{EV}. \chadded{Nevertheless, further research on the effects of higher charging powers on the standard measurement is encouraged to reduce the overall testing duration.}

Furthermore, \ac{EV}-specific peculiarities that influence the proposed procedure must be considered, which are discussed below.

\begin{figure}[h]
	\centering
	\makebox[0pt]{\includegraphics[width=17cm]{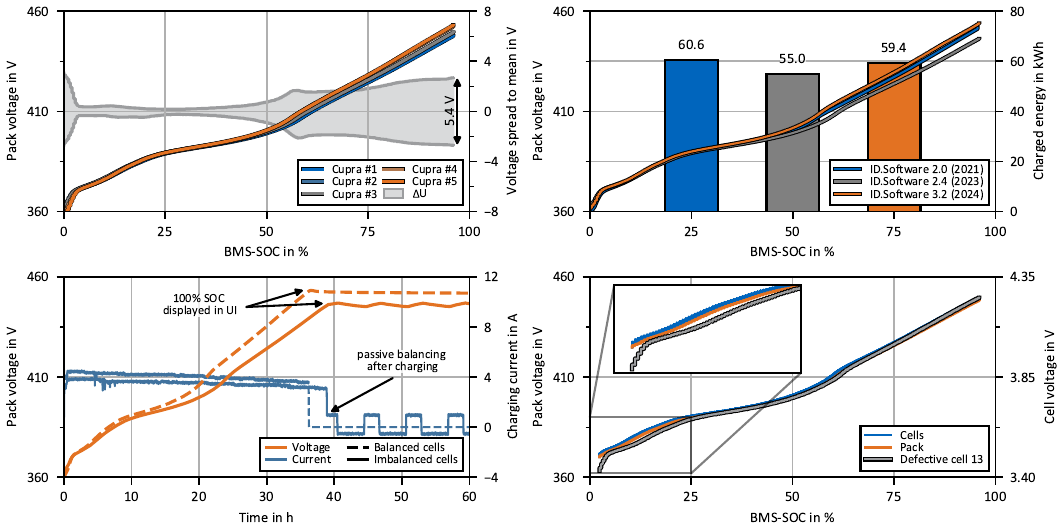}}
	\caption{Peculiarities of \acp{EV} and their influence on the proposed standard measurement. a) Measuring a fleet of identical vehicle models with comparable distance covered shows voltage differences so that the vehicle's individual capacity/energy content is not reflected with a fixed voltage window. b) Software updates can change \ac{BMS} settings, such as changing the voltage operation window as shown for annual measurements of the exact vehicle. c) Balancing occurs when cells within the pack have imbalanced \ac{SOC} levels. Balancing phases after reaching \SI{100}{\percent} UI-\ac{SOC} are triggered by the \ac{BMS} to realign the cells' voltages. However, \acp{EV} in an unbalanced state can lead to inconsistent cut-off voltages. The consecutive measurements were conducted on the exact vehicle. d) Defective cells influence the battery system, e.g., cut-off conditions are met prematurely, and the measured capacity/energy is reduced. Note that for these \ac{VW} vehicles, \SI{96}{\percent} is the maximum \ac{BMS}-\ac{SOC}, which equals \SI{100}{\percent} UI-\ac{SOC}.}
	\label{fig:cons}
\end{figure}

For instance, measuring a fleet of the same vehicle model at similar mileages, shown in \autoref{fig:cons}a for five Cupra Borns, exhibits deviations in the pack voltages. Although cell-to-cell variations are of small order, they can become significant in the cell assembly~\cite{Rumpf.2017}. Furthermore, individual onboard \ac{SOC} estimations refer to different voltages, i.e., when all Cupras were completely charged, showing \num{100}\,\si{\percent}~\ac{SOC} in the \ac{UI}, the voltages differed up to \SI{5.4}{\volt}. Therefore, a fixed voltage window must be carefully defined to fit all vehicles of a model. This, however, affects the determination of the \ac{EV}'s individual energy and capacity, as the fixed voltage ranges for the standard measurement crop the actual usable voltage window available in the vehicle. To compensate for this, each vehicle would need individual cut-off voltages, thereby inhibiting comparability. \chadded{Differences in charging between fixed \ac{SOC} and voltage boundaries, as well as the repeated measurements of one specimen, are given in \autoref{tab:cupra_aging}, showcasing that fixing a voltage range is better suited for the comparability and reproducibility of the measurement results.}

Likewise, interference of the \ac{BMS} can hamper the standard measurement, e.g., software updates, possibly over-the-air, can change the voltage range of a battery pack, altering the usable operation window and capacity/energy retention, as shown in \autoref{fig:cons}b. Therefore, changes in the battery system must transparently be communicated to the vehicle owner so that the standard measurement and the reference for the \ac{SOH} can be adjusted. Moreover, balancing through the \ac{BMS} takes place to align the voltage levels of cells in the pack that have drifted due to (increasingly) differing aging levels, as seen in \autoref{fig:cons}c. Measuring a vehicle in an unbalanced state influences the standard measurement as the upper voltage limit might not be consistent, and features in \ac{DV} curves can fade~\cite{Reiter.2023}. Therefore, the cells in the pack are ideally balanced prior to the measurement. 

Also, a cell defect within the battery pack~(\autoref{fig:cons}d) determines the usable capacity/energy, as cut-off voltages are met prematurely~\cite{Baumann.2018}. As a result, the defined voltage range may not be reachable on the pack level, and the \ac{SOH} cannot be calculated with the proposed standard measurement. Nonetheless, if cell-level voltage data is available, targeted maintenance is possible to detect and locate faulty cells.

Regarding degradation assessment by \ac{DV} analysis on the vehicle level, the previous examples have shown that degradation modes cannot always be determined, and limitations exist, e.g., \ac{BMS} interference~\cite{Bilfinger.2024}.
Also, it is unlikely for \acp{OEM} to publish voltage-capacity curves of pristine specimens for the evaluation of degradation modes. However, test facilities can establish databases for assessing aging between \acp{EV} of the same model type, as shown for the Porsche Taycan and Tesla Model Y.

Despite the limitations discussed, we are convinced that the proposed standard measurement based on charging is useful for the \ac{SOH} determination for \ac{EV}s' battery packs, and additional information from \ac{DV} analysis adds value through deeper battery aging assessment. Thus an independent, external, and transparent battery diagnosis for \acp{EV} is enabled. \chadded{Further, to maximize its effectiveness, we recommend that EV owners perform the proposed measurement at the beginning of the vehicle’s life to establish a reference point for future evaluations, and enabling the possibility for individual aging analysis.}



Nonetheless, due to the complexity of an \ac{EV}'s battery pack, a capacity- and energy-based \ac{SOH} are merely two of many parameters necessary to fully characterize its multi-dimensional \ac{SOH}~\cite{Bulow.2024}. For instance, we also encourage a standard measurement for the internal resistance ($SOH_R$) as it increases throughout an \ac{EV}'s lifetime and influences the usable \ac{SOC} range, especially under high-power scenarios.

Hereby, the communications norms \ac{OCPP}~2.1 and \mbox{ISO15118-20} can be useful, enabling bidirectional charging where more sophisticated battery diagnosis is possible through tailored (dis-)charging procedures, e.g., incorporating power pulses~\cite{Willrett.2023}. It is assumed that future \acp{EV} will comply with these norms and are capable of bidirectional charging.

Furthermore, data from vehicle sensors must be available for any onboard battery diagnostic method, e.g., through the \ac{OBD} interface, without encryption in sufficient resolution, accuracy, precision, and sample rate, at least for the voltage, current, and \ac{SOC} on the pack level. Additional access to further information, such as module temperatures and the cell voltage across each series connection, enables proactive and tailored maintenance, which is also encouraged. Thus, the \acp{OEM} must provide the required data, reference values, and the possibility of external battery diagnosis. For instance, Tesla's 'battery health' implementation is a step in the right direction but lacks transparency about the specific measurement procedure and how the \ac{SOH} is calculated~\cite{batteryhealth_tesla}.

Ultimately, the legislator must ensure that independent battery diagnosis is possible, e.g., through the CEN/CENELEC standardization request M/579 aiming at "performance, safety and sustainability requirements for batteries", where our proposal can suit as a discussion basis~\cite{ec2024transparency}.

\section*{Methods}
\label{ch:methods}

The \acp{EV} under study are presented in \autoref{sec:vehicles}. Further, the measurement procedure~(\autoref{sec:procedure}), vehicle data acquisition~(\autoref{sec:data_ac}), and postprocessing~(\autoref{sec:post}) are explained in closer detail. 

\subsection*{Vehicles under test}
\label{sec:vehicles}

The vehicles under test are state-of-the-art \acp{EV} from \ac{VW}, Cupra, Porsche and Tesla. The vehicles' battery specifications are summarized in \autoref{tab:vehicles}. All vehicles were purchased from a dealership or leased from non-\ac{OEM} research partners to ensure unmodified mass-production specimens and an uninfluenced examination.

Seven Volkswagen Group vehicles sharing the same battery pack with \ac{NMC}532/graphite cells were tested: Five Cupra Borns with similar mileages~(\SIrange{12700}{17700}{\kilo\meter}) and two \ac{VW} ID.3s. Both first-generation \ac{VW} ID.3s were purchased in 2020. The vehicle analyzed for aging was comprehensively investigated through teardown analysis by Wassiliadis et al.~\cite{WASSILIADIS2022100167}. This specimen was subject to multiple measurements between mileages of \SI{400}{\kilo\meter} and \SI{40500}{\kilo\meter} at different software releases and balancing states and was previously analyzed by \ac{DV} analysis at lower mileages~\cite{Bilfinger.2024}. The second \ac{VW} ID.3 has a defective cell limiting the overall battery pack performance.

The 2020 Tesla Model 3 Standard Range Plus is equipped with prismatic \ac{LFP}/graphite cells that have been comprehensively investigated at the cell level~\cite{Stock.2023} and vehicle level~\cite{rosenberger2024quantifying}. The vehicle was measured at mileages \SI{1000}{\kilo\meter} and \SI{80100}{\kilo\meter} and was previously analyzed by \ac{DV} analysis at lower mileages~\cite{Bilfinger.2024}.

Two 2022 Porsche Taycans from different sources were measured at mileages of \SI{10000}{\kilo\meter} and \SI{70500}{\kilo\meter}. Both vehicles share the same battery pack with \ac{NMC}622/graphite cells. Also two Tesla Model Y Long Range vehicles~(manufactured in 2022 and 2024) with mileages of \SI{77600}{\kilo\meter} and \SI{24800}{\kilo\meter} were measured. Ank et al.~\cite{Ank.2023b} investigated the \ac{NMC}~811/graphite chemistry used by both vehicles. These measurements show that aging assessment through the proposed standard measurement is comparable within vehicle models.

\subsection*{Measurement procedure}
\label{sec:procedure}

A battery cell's remaining capacity/energy is typically determined during reoccurring check-ups from controlled (dis-)charging cycles between fixed boundary constraints~\cite{Severson.2019}. This general procedure is transferred onto the vehicle level through a controlled charging cycle adapted to the vehicles' usage window, similar to Wassiliadis et al.~\cite{WASSILIADIS2022100167}. Note that the \ac{SOC} displayed in the \ac{UI} differs from the \ac{BMS}-\ac{SOC}, possibly to preserve energy buffers. Here, only the  \ac{BMS}-\ac{SOC} is considered.

The vehicles are fully discharged to \SI{0}{\percent} \ac{SOC} before the measurement, whereby the last few percent are drained using auxiliary units. 
After the discharging procedure, the vehicles were set to rest for approx. \SI{30}{\minute} for further overpotentials to settle before the charging session is triggered by connecting the vehicle to the \ac{AC}~(IEC~61851 mode) charging device Juice Booster 2~(Juice Technology, Switzerland). The charging device can supply charging powers between \SI{1.4}{\kw} and \SI{22}{\kw} depending on the connected socket~(single/triple phase). A quasi-stationary measurement is sought by preferring low charging powers to reduce the effects of overpotentials and inhomogeneity in the \ac{NE}~\cite{Lewerenz.2018}. Therefore, the charging power is set to yield at least a \SI{30}{\hour} charging session calculated by the vehicles' net energy content. Other authors found that a \num{15}\,\si{\hour} charging session is sufficient and does not mitigate the \ac{SOH} calculation or aging assessment~\cite{Schmitt.2023}.
The vehicles are exclusively charged inside a workshop at an ambient temperature of \SI{20}{\celsius} until the \ac{BMS} terminates the charging session. The capacity and energy are calculated between fixed voltage ranges, given in \autoref{tab:vehicles} defined for each \ac{EV} model.

\subsection*{Data acquisition}
\label{sec:data_ac}

The measurement procedure relies on onboard sensor data of the vehicles under test, which is recorded using different methods depending on the manufacturer. The most relevant \ac{HV} battery pack level signals are the voltage, current, and \ac{BMS}/\ac{UI}-\ac{SOC}. The cell voltages across each serial connection and temperature data are also sampled when available. The data is recorded with the precision of the built-in sensors that have sufficient data quality for further processing~\cite{Bilfinger.2024}.

The entire \ac{CAN} bus communication is recorded for the Tesla vehicles via a \ac{CAN} interface with the recording tool Busmaster. The data is converted into physical and human-readable quantities by a publicly available .dbc file from {\href{https://github.com/joshwardell/model3dbc}{{GitHub}}}. The correctness of the conversion schemes is asserted by comparison to data from a commercially available vehicle diagnostic device, which is unsuitable for logging over a long period. The sample rate is \SI{100}{\hertz} for the pack current, voltage, and \ac{SOC} and \SI{0.5}{\hertz} for the cell voltages. The resolution of the digitized signals can be estimated from the sampled data as \SI{0.1}{\volt} for the voltages, \SI{0.1}{\ampere} for the current, and \SI{0.1}{\percent} for the \ac{SOC}.

Data from the \ac{VW} group vehicles is queried by \ac{UDS} sent through the \ac{OBD}-II diagnostic interface and converted with a custom {\href{https://github.com/TUMFTM/uds-decoder/}{{UDS decoder}}} tool. Each request addresses a unique \ac{ID} reverse-engineered by reproducing the traffic from a commercially available vehicle diagnostic device. An in-depth description of the data acquisition is given by Merkle et al.~\cite{Merkle.2021}. During the measurement, the individual values are requested at \SI{1}{\hertz}. However, the gateway board, responsible for handling requests, responds to requests depending on the available bandwidth. In practice, a sample is logged in a frequency between \SI{0.1}{\hertz} and \SI{1}{\hertz}. The resolution of the digitized signals can be estimated from the sampled data as \SI{0.25}{\volt} for the voltages, \SI{0.1}{\ampere} for the current, and \SI{0.40}{\percent} for the \ac{SOC}.

\subsection*{Data postprocessing}
\label{sec:post}

The long-format data traces are subsequently synchronized onto a single time vector by linear interpolation and cropped to fit the voltage range specified in \autoref{tab:vehicles}. The data is filtered by a mean filter in the forward direction with a window size of \SI{1}{\percent} of data points to generalize between different vector lengths. The capacity $Q_{calc}$ and $E_{calc}$ are calculated by \autoref{eq:capacity} and \autoref{eq:energy}, respectively.

The \ac{DV} curve is calculated from the derivatives of the pack voltage and capacity signals by
\begin{align}
	DV = \frac{\nicefrac{dU}{dt}}{\nicefrac{dQ}{dt}} \approx \frac{\Delta U}{\Delta Q}.
	\label{eq:DV}
\end{align}

For practical applications, the \ac{DV} curve is calculated by forward difference schemes~\cite{Smith.2012}. Furthermore, the \ac{DV} curve is normalized by the nominal capacity $Q_N$ given in \autoref{tab:vehicles}, through which the data is scaled in y-direction around the number of series-connected cells in the battery pack, enhancing the comparability between battery systems with different capacities. Care must be taken when applying filtering methods, as these can alter features of interest~\cite{Dubarry.2022}.

\backmatter
\section*{Declarations}
\bmhead{Data availability}We want to give any researcher access to our data without any limits. The data of all measurements can be found at \href{https://mediatum.ub.tum.de/1765567}{mediaTUM}. Press login at top left of page. Access: reviewer-access-12, password: QjvP@Lv-ea848!22hUkPM
\bmhead{Code availability}The code for processing the data and generating the figures is available at \href{https://github.com/TUMFTM/nature_soh.git}{FTM Github}.
\bmhead{Acknowledgements}
This research received funding from the Bavarian Ministry of Economic Affairs, Regional Development, and Energy within the project 'charge.COM' under the grant number DIK-0262/02. Furthermore, this work was partly funded by the German Federal Ministry for Economic Affairs and Climate Action~(BMWK) within the project 'ultraBatt' under the grant number 01MV21015D and the German Federal Ministry of Education and Research~(BMBF) within the project ’BALd’ under grant number 03XP0320B. The funders played no role in the study design, data collection, analysis, and interpretation of data or the writing of this manuscript. We also thank TUM.Battery for promoting battery research by providing an interdisciplinary network of experts.
\bmhead{Author contributions}
P.B. initiated the idea, conducted measurements, and wrote the paper. M.S. conducted measurements, contributed to the standard measurement, and interpretation of the results. P.R. conducted measurements, contributed to the methods and discussion. K.A.G., J.S., and C.G. supported the measurements, analyzed the data, and addressed customer relevance. M.L. contributed to the concept and supervised the research. All authors discussed the results, wrote sections, and revised the paper thoroughly and critically.
\bmhead{Conflict of interest/Competing interests}The authors declare that they have no known competing financial interests or personal relationships that could have appeared to influence the work reported in this paper.
\noindent


\bibliography{main.bib}

\begin{minipage}{\columnwidth}
	\footnotesize
	\captionof{table}{Overview of vehicle specifications from the vehicles under study.}
	\label{tab:vehicles}
	\vspace{0pt}
	\addtolength{\tabcolsep}{-0.15em}
	\centering
	\begin{tabularx}{\columnwidth}{lccccc}
		\toprule
		Attribute& Unit & \makecell{VW ID.3\\ Pro Performance\textsuperscript{e}}&Porsche Taycan&\makecell{Tesla Model 3\\ SR+ (LFP)}&\makecell{Tesla Model Y\\ LR (NMC)}\\
		\midrule
		Gross energy\textsuperscript{a}&\si{\kwh}&62&93.4&55&77\\ 
		Net energy\textsuperscript{b}&\si{\kwh}&58&82.3&52.5&75\\ 
		Nom. capacity&\si{\ah}&145\textsuperscript{d}&112\textsuperscript{d}&161.5\textsuperscript{c}&211.6\textsuperscript{c}\\ 
		Voltage range\textsuperscript{d}&\si{\volt}&360 -- 450&650 -- 830&335 -- 365&300 -- 400\\
		Serial cells\textsuperscript{b}&-&108&198\textsuperscript{c}&106&96\\
		Parallel cells\textsuperscript{b}&-&2&2\textsuperscript{c}&1&46\\
		Cell format\textsuperscript{b}&-&pouch&pouch\textsuperscript{c}&prismatic&cylindrical\\ 
		Chemistry\textsuperscript{c}&-&C/NMC532&C/NMC622&C/LFP&C/NMC811\\ 
		\bottomrule\\[-6pt]
		\multicolumn{6}{l}{\textsuperscript{a} Vehicle registration documents}\\
		\multicolumn{6}{l}{\textsuperscript{b} EV Database: {\href{https://ev-database.org/de/pkw/1831/Volkswagen-ID3-Pro}{{Volkswagen ID.3 Pro Performance}}, \href{https://ev-database.org/de/pkw/2098/Porsche-Taycan}{{Porsche Taycan}}, }}\\
		\multicolumn{6}{l}{\href{https://ev-database.org/de/pkw/1320/Tesla-Model-3-Standard-Plus-LFP}{{\hspace{2.05cm}Tesla Model 3 Standard Range Plus}}, {\href{https://ev-database.org/de/pkw/1619/Tesla-Model-Y-Maximale-Reichweite-AWD}{{Tesla Model Y Long Range}}}}\\
		\multicolumn{6}{l}{\textsuperscript{c} Literature: \ac{VW}~\cite{F.J.Gunter.2022},  {\href{https://www.heise.de/news/Elektroauto-Porsche-Taycan-Modellpflege-bringt-hoehere-Effizienz-mit-neuem-Akku-9621738.html}{{Porsche Taycan}}}, Tesla Model 3~\cite{Stock.2023}, Tesla Model Y~\cite{Ank.2023b}}\\
		\multicolumn{6}{l}{\textsuperscript{d} Measurement data}\\
		\multicolumn{6}{l}{\textsuperscript{e} Same specifications for Cupra Born}\\
	\end{tabularx}
\end{minipage}
\begin{minipage}{\columnwidth}
	\footnotesize
	\captionof{table}{Comparison of charged energy during the standard measurement across five Cupras with fixed SOC and voltage windows. Cupra 213 was measured three times to assess reproducibility.}
	\label{tab:cupra_aging}
	\vspace{0pt}
	\addtolength{\tabcolsep}{-0.25em}
	\centering
	\begin{tabularx}{\columnwidth}{lC{1.5cm}|C{2cm}C{2cm}|C{1.5cm}C{2cm}}
		\toprule
		\multicolumn{2}{l}{Measuring window}  & \multicolumn{2}{c}{\SIrange{0}{100}{\percent}~SOC} & \multicolumn{2}{c}{\SIrange{370}{445}{\volt}\textsuperscript{b}}\\
		\midrule
		Vehicle\textsuperscript{a}& Mileage in km& Voltage range in \si{\volt} & Charged energy in \si{\kwh} & SOC range in \si{\percent} & Charged energy in \si{\kwh}\\
		\midrule
		Cupra 397 & 12750 &359.0 -- 447.6 & 56.5 & 3.2 -- 93.6 & 53.3 \\
		Cupra 349 & 12600 &363.0 -- 449.3 & 57.8 & 2.4 -- 92.4 & 53.9 \\
		Cupra 288 &13250 &361.8 -- 449.3 & 57.7 & 2.4 -- 92.0 & 53.7 \\
		Cupra 204 &17700 &358.8 -- 452.2 & 58.7 & 3.2 -- 89.6 & 52.7 \\
		Cupra 213~(I) &15300 &357.2 -- 453.0 & 60.5 & 3.2 -- 89.2 & 53.8 \\[1pt] \midrule \midrule
		Cupra 213~(II) &15300 &359.0 -- 453.0 & 60.5 & 3.2 -- 89.2 &  53.9 \\
		Cupra 213~(III) &15300 &363.5 -- 452.6  & 59.9 & 2.4 -- 89.2 & 54.0 \\
		\bottomrule\\[-6pt]
		\multicolumn{3}{l}{\textsuperscript{a} The number refers to the vehicle's licence plate.}\\
		\multicolumn{3}{l}{\textsuperscript{b} The voltage range was chosen to fit all specimens.}
	\end{tabularx}
\end{minipage}
\end{document}